\documentclass[11pt,letterpaper]{article}
\usepackage{systeme} %para usar sistemas de ecuaciones
\usepackage[utf8]{inputenc}
\usepackage{multirow} %para las tablas
\usepackage{booktabs} %para la estética de tablas
\usepackage[english]{babel}
\usepackage{amsmath}
\usepackage{amsfonts}
\usepackage{amssymb}
\usepackage{graphicx}
\usepackage[small,bf]{caption} %Para caption con letra pequeña y Figura en negrita
\usepackage[left=3cm,right=3cm,top=2cm,bottom=3cm]{geometry}

\author{Jorge Pinochet}
\title{\textbf{Classical Tests of General Relativity Part II: Looking to the Past to Understand the Present}}
\begin{document}

\author{Jorge Pinochet$^{*}$\\ \\
 \small{$^{*}$\textit{Departamento de Física, Universidad Metropolitana de Ciencias de la Educación,}}\\
 \small{\textit{Av. José Pedro Alessandri 774, Ñuñoa, Santiago, Chile.}}\\
 \small{e-mail: jorge.pinochet@umce.cl}\\}

\date{}
\maketitle

\begin{center}\rule{0.9\textwidth}{0.1mm} \end{center}
\begin{abstract}
\noindent The objective of this second part of the work is to present heuristic derivations of the three classical tests of general relativity. These derivations are based on the Einstein equivalence principle and use Newtonian physics as a theoretical framework. The results obtained are close to Einstein's original predictions. Historical and anecdotal aspects of the subject are also discussed. \\ \\


\begin{center}\rule{0.9\textwidth}{0.1mm} \end{center}
\end{abstract}

\maketitle

\section{Introduction}
The main objective of this second part of the work is to present heuristic derivations of the three classical tests of general relativity (GR). The first derivation is a variant of an argument presented by the author in another work [1], the second derivation is original, and the third is a variant of an argument that appears in some texts of modern physics [2]. These derivations are based on the Einstein equivalence principle and use Newtonian physics as a theoretical framework. Despite these simplifications, the results obtained hardly differ from Einstein's original predictions. Although we will not delve into the notion of space-time curvature, it is important to keep in mind that a detailed explanation of the classical tests requires the notion of curvature, as discussed shallowly in Part I.\\

The article is organised as follows. We first analyse the deflection of light by the Sun. We will then examine the perihelion precession of Mercury and finally tackle the gravitational redshift of light. Although this sequence does not coincide with the historical order of events (which is not linear), in the author's opinion, it is the most convenient order from a pedagogical point of view. The article ends with some comments on the current and future scientific status of GR.

\section{Deflection of light by the Sun}

The basic idea behind this classical test is illustrated in Fig. 1. A ray of light that passes very close to the surface of a massive celestial body, such as the Sun, suffers a deflection from its straight path (this deflection is a manifestation of the curvature of spacetime around the Sun). If the light comes from a distant star, it will show an apparent position different from its actual position.\\

From the Einstein equivalence principle, it is possible to predict the deflection of light by the Sun and derive the approximate equation that describes this effect. To achieve this, we can consider a spacecraft moving with constant acceleration $g$ in a region of the universe without gravity. A photon enters through a window on the left side wall and reaches the other end of the spacecraft. Fig. 2 shows two perspectives of the photon (drawn as a red dot) at four equidistant instants of time. The left image shows the point of view of an observer who is outside the spacecraft, in the reference system where the photon is emitted. For this observer, the photon follows a straight path. The right image shows the point of view of an observer located inside the spacecraft. For this observer the photon "falls", describing a parabolic path.\\

As Fig. 2 shows, during its journey within the spacecraft, the photon travels a horizontal distance $l$ and descends or falls a vertical distance $s$. For an observer inside the spacecraft, the distance the photon falls is:

%eq 1
\begin{equation}
s = \frac{1}{2} gt^{2}.
\end{equation}

We can estimate the time it takes for the photon to fall this distance as:

%eq 2
\begin{equation}
t = \frac{l}{c},
\end{equation}

where $c = 3 \times 10^{8} m\cdot s^{-1}$ is the speed of light in vacuum. By introducing this result in Eq. (1), we obtain:

%eq 3
\begin{equation}
s = \frac{gl^{2}}{2c^{2}}.
\end{equation}

We will use this equation again when we discuss the perihelion precession of Mercury in the next section. Due to the enormous value of $c$, in general, the angle of deviation will be extremely small. Said angle (in radians) can be obtained by deriving Eq. (3):

%eq 4
\begin{equation}
\alpha \cong \frac{ds}{dl} = \frac{gl}{c^{2}}.
\end{equation}

\begin{figure}
  \centering
    \includegraphics[width=0.7\textwidth]{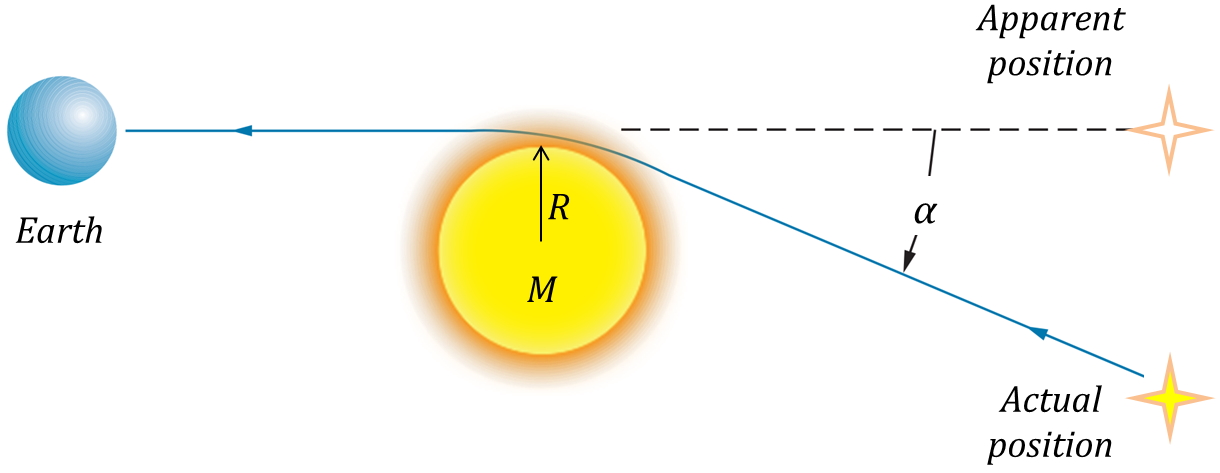}
  \caption{Deflection of light by the Sun. An observer on Earth perceives the star to be in an apparent position other than the actual one.}
\end{figure}

By virtue of the Einstein equivalence principle, we conclude that if the photon undergoes a deviation $\alpha$ within the spacecraft with acceleration $g$, it will undergo the same deviation if the spacecraft is on the surface of the Sun, where the acceleration of gravity is $–g$ (see Fig. 3). Then, Eq. (4) must be valid in the vicinity of the Sun (and of any spherical celestial body) and we can express $g$ in the form:

%eq 5
\begin{equation}
g = \frac{GM}{R^{2}},
\end{equation}

where $M$ is the mass of the Sun and $R$ is its radius. By eliminating g between Eqs. (4) and (5), we get:

\begin{equation}
\alpha \cong \frac{GMl}{c^{2} R^{2}}.
\end{equation}

\begin{figure}
  \centering
    \includegraphics[width=0.7\textwidth]{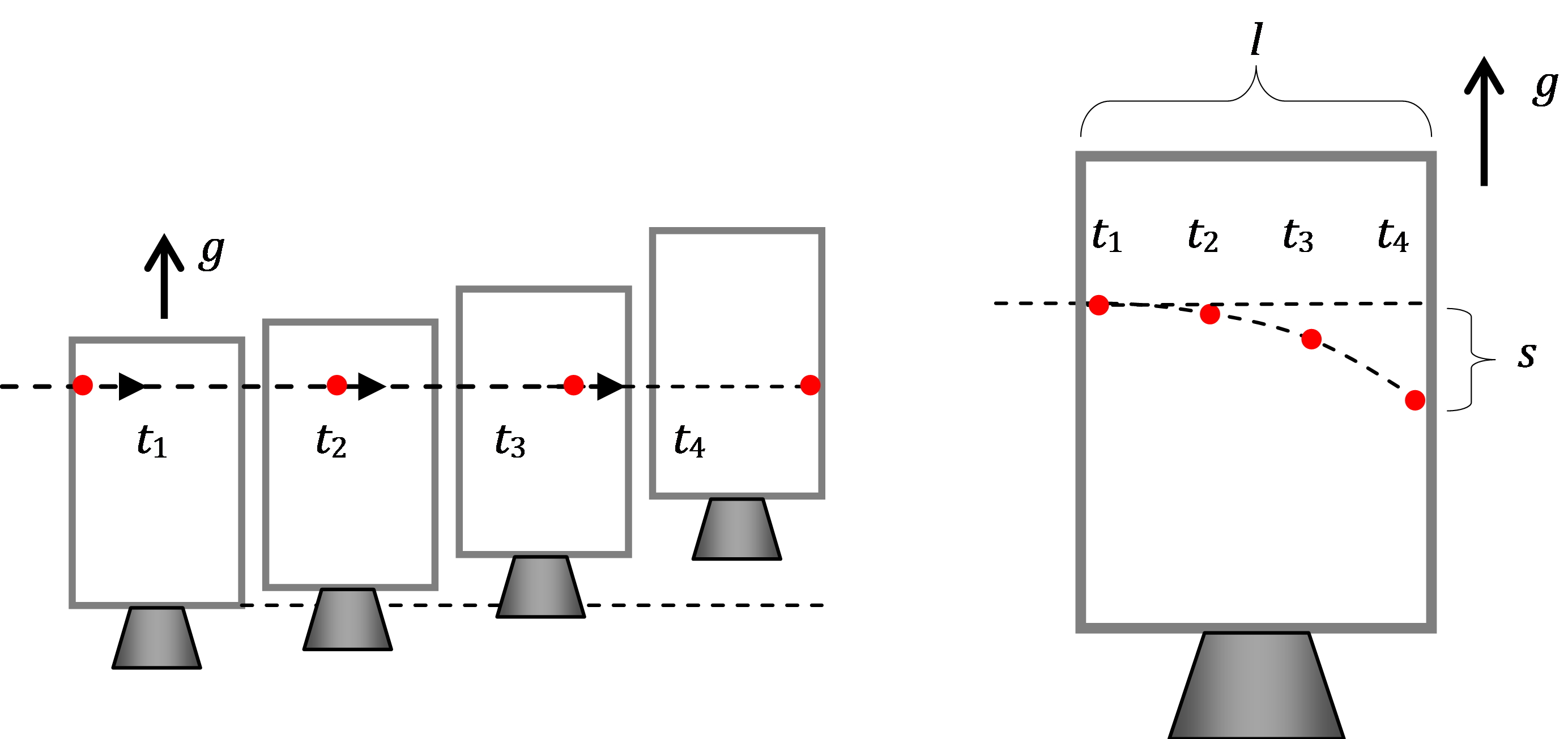}
  \caption{Left: Path of the photon as seen by an inertial observer in four instants of time. Right: Path of the photon seen from inside the spacecraft, in the same instants of time.}
\end{figure}

Although, strictly speaking, Eq. (6) only has local validity, that is, it is only applicable in a small region on the solar surface, where $g$ is uniform, we can estimate the non-local deflection of a photon that skims the surface of the Sun assuming that the spacecraft is wide enough so that $l = 2R$:

%eq 7
\begin{equation}
\alpha \cong \frac{2GM}{c^{2} R}.
\end{equation}

The equation found by Einstein is: 

%eq 7
\begin{equation}
\alpha_{E} = \frac{4GM}{c^{2} R}.
\end{equation}

We see that  $\alpha_{E} = 2\alpha$, so that Eq. (7) is a good approximation, especially considering the simplicity of the derivation. Eq. (8) was first verified by two English astronomical expeditions conducted during the total eclipse of the Sun on May 29, 1919. The expeditions sought to determine the deflection caused by the Sun over the light of stars located behind the solar disk (the stars were visible thanks to the eclipse). One of the expeditions was led by Arthur Eddington and Frank Dyson, and was conducted on Prince Island, off the east coast of Africa. The other expedition was led by Andrew Crommelin and Charles Davidson and was conducted in Sobral, Brazil. The plan was to compare images of the stars taken during the eclipse (apparent position) with images of the same stars taken six months later (actual position), when the Sun does not come between the Earth and stars.\\

\begin{figure}
  \centering
    \includegraphics[width=0.3\textwidth]{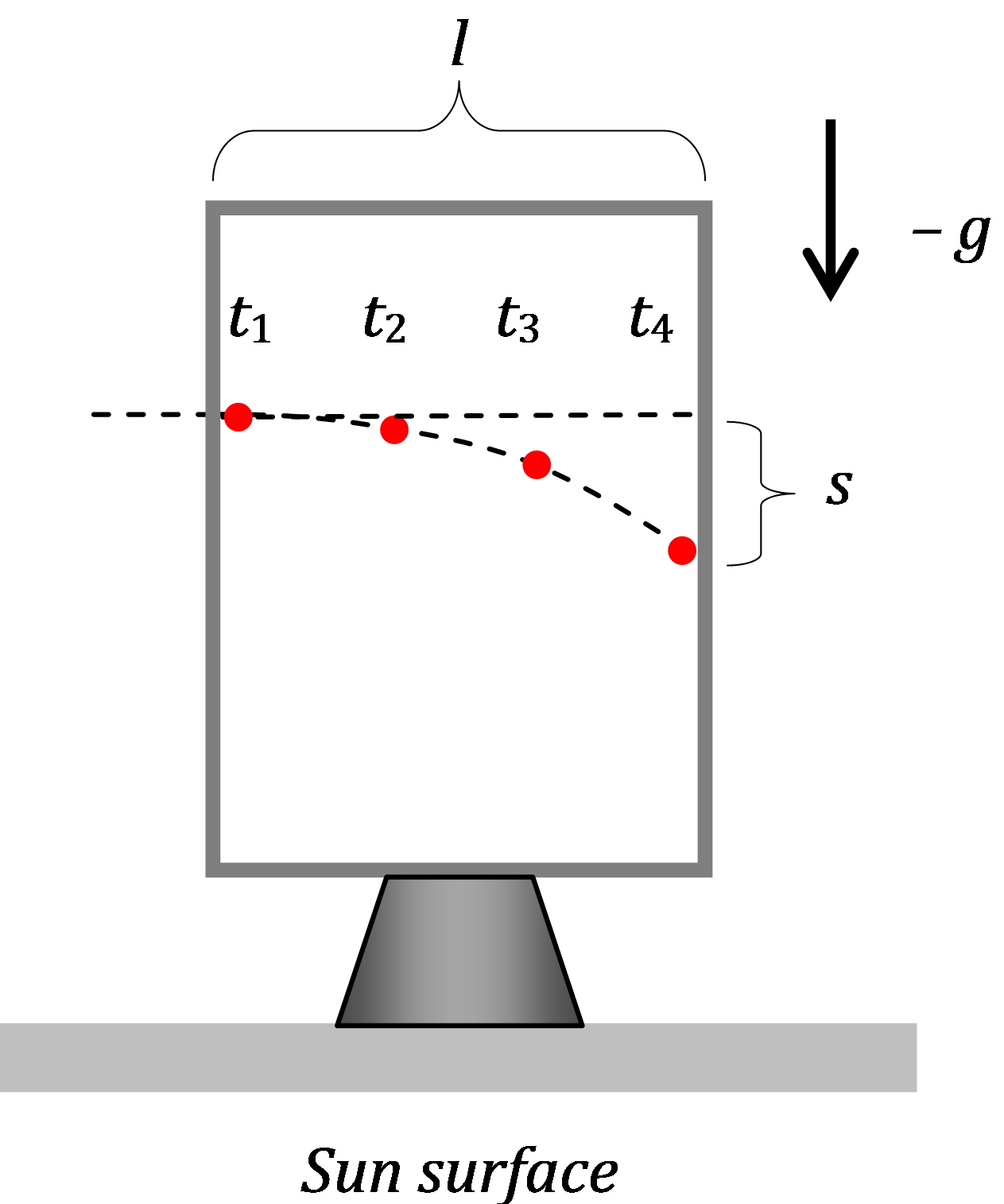}
  \caption{Spacecraft at rest on the surface of the Sun. A photon that enters through the lateral window experiences a deflection similar to that illustrated in Fig. 2, right.}
\end{figure}

If in Eq. (8) we consider $M = 1.99 \times 10^{30} kg$ (solar mass) and $R = 6.95 \times 10^{8} m$ (solar radius), and we introduce the other constants, we obtain $\alpha = 4.244 \times 10^{-6} rad$. As $1 rad = 360^{o} /2\pi$ and $1^{o} = 3600''$, we get:

%eq 9
\begin{equation}
\alpha = 4.244 \times 10^{-6} \times \frac{360^{o}}{2\pi} \times \frac{3600''}{1^{o}} = 1.75''.
\end{equation}

Within observational uncertainties, the value found by astronomical expeditions coincides with Eq. (9) [3]. Newton's law of gravitation does not predict any deflection for light in a gravitational field\footnote{If we assume that light is a material particle beam, as Newton believed, the deviation of light can be calculated, obtaining a figure that is half the value calculated in Eq. (8), and therefore coincides with Eq. (7). However, we know that light is not made up of material particles.}, so the results obtained by the expeditions not only meant a triumph for GR, but also the decline of the Newtonian worldview.

\section{Perihelion precession of Mercury}

According to the Newton-Kepler laws, an isolated planet that revolves around a star, such as the Sun, will describe an elliptical orbit with the star in one of the foci. This means that the angle described by the radius vector (the line connecting the planet with the star) between one perihelion and the next is zero. That is, after completing one lap, the perihelion returns to the starting point.\\

However, when calculations are made using GR, a perihelion precession or perihelion advance of the planet is found, which means that between one perihelion and the next, the perihelion does not return to its initial point, and consequently the angle described by the radio vector is slightly greater than zero, as illustrated in Fig. 4. The closer a planet is to its star, the greater this effect is. Einstein applied GR to calculate the perihelion advance of Mercury, the planet closest to the Sun.\\

\begin{figure}
  \centering
    \includegraphics[width=0.7\textwidth]{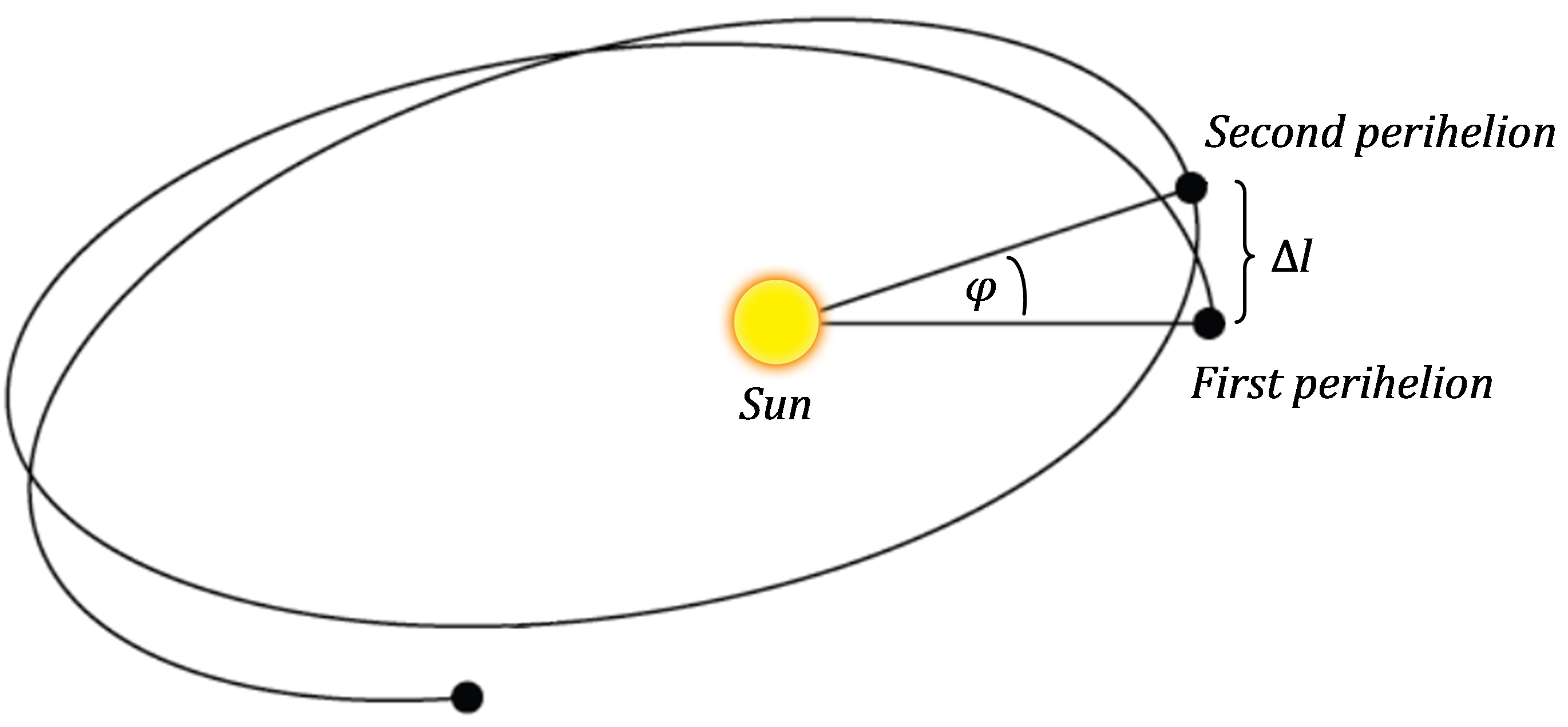}
  \caption{Mercury (black circle) in two consecutive positions of its perihelion. For each revolution, a perihelion advance occurs at an angle $\varphi$, which corresponds to an additional orbital displacement $\Delta l$.}
\end{figure}

We can make an approximate calculation of the perihelion advance of Mercury using the Einstein equivalence principle. However, in this case we will use it implicitly, drawing on the ideas developed in the previous section. To do this, suppose that Eq. (3) allows us to describe both the angular deviation of a photon and the angular deviation of Mercury (or of another small celestial body in relation to the Sun) in the solar gravitational field. Suppose also that Mercury is the only planet that revolves around the Sun and that its elliptical orbit does not differ much from a circumference.\\

By applying Eq. (3) to the perihelion advance, we see that $l$ plays the role of a circumference of perimeter $2\pi r$, where $r$ is the radius of Mercury’s orbit. From Eq. (3), we find that in each revolution, Mercury advances (deviates angularly) between one perihelion and the next an approximate distance:

%eq 10
\begin{equation}
s \cong \frac{(2\pi r)^{2}}{2c^{2}} g = \frac{2\pi^{2}}{c^{2}} GM,
\end{equation}

where we introduce the value of $g$ given by Eq. (5). Furthermore, Kepler's third law for circumferential orbits states that:

%eq 11
\begin{equation}
\frac{GM}{4\pi^{2}} = \frac{r^{3}}{T^{2}},
\end{equation}

where $T$ is the Mercury orbital period and $r$ is the radius of its orbit. By eliminating $GM$ between Eqs. (10) and (11), we get:

%eq 12
\begin{equation}
s \cong \frac{8\pi^{4} r^{3}}{c^{2} T^{2}}.
\end{equation}

Dividing by $r$, we obtain the perihelion advance of Mercury per revolution, expressed in radians (see Fig. 4):

%eq 13
\begin{equation}
\varphi \cong \frac{s}{r} = \frac{8\pi^{4} r^{2}}{c^{2} T^{2}}.
\end{equation}

The equation originally found by Einstein is:

%eq 14
\begin{equation}
\varphi_{E} = \frac{24\pi^{3} a^{2}}{c^{2} T^{2} (1- e^{2})},
\end{equation}

where $e$ is the eccentricity of the elliptical orbit and $a$ is the semi-major axis. For $r \approx a$ and $e \ll 1$ at Eq. (14), we see that $\varphi / \varphi_{E} = \pi /3 \cong 1.05$, meaning that Eq. (13) is an excellent approximation.\\

Since the mid-19th century, astronomers have observed a perihelion advance of Mercury of $57''$ per century, which they attributed to the gravitational influence of the rest of the planets. By applying Newton's law of gravitation, it was possible to predict an advance of $531''$ per century, so that there were $574'' – 531' = 43''$ per century without explanation [3]. Using GR, Einstein was able to explain this difference. Indeed, in the case of Mercury, we know that $a = 57.9 \times 10^{9} m$, $T = 10^{7} s$ and $e = 0.206$. Considering these values in Eq. (14), and introducing the other constants, we obtain $\varphi_{E} = 5 \times 10^{-7} rad/revolution$, or:

%eq 15
\begin{equation}
\varphi_{E} = \left( 5 \times 10^{-7} rad \times \frac{360^{o}}{2\pi rad} \times \frac{3600''}{1^{o}} \right) / revolution = 0.103''/revolution.
\end{equation}

This is the perihelion advance of Mercury without considering the gravitational influence of the other planets, that is, subtracting said influence. Since Mercury makes 415 revolutions in a century, we conclude that the perihelion advance is $415 \times 0.103 \cong 43''$ per century, which is exactly the advance that Newtonian gravitation cannot explain. This figure was the first empirical confirmation of GR.

\section{Gravitational redshift of light\protect\footnote{Strictly speaking, the gravitational redshift of light is a test of the Einstein equivalence principle, while the other two are tests of the GR (Schwarzschild solution) in the weak field limit.}}

According to GR, the light that moves away from a celestial body, like the Earth, experiences a decrease in its energy, so that a distant observer detects the light with a redshift, that is, the wavelength presents a systematic shift towards the low frequency range. This is similar to what happens to an object that is thrown up from the Earth's surface. As the object rises, moving against gravity, it loses speed and kinetic energy. However, since light always moves with constant speed $c$, the only way that energy loss can occur is through an increase in wavelength or a decrease in frequency.\\

\begin{figure}
  \centering
    \includegraphics[width=0.2\textwidth]{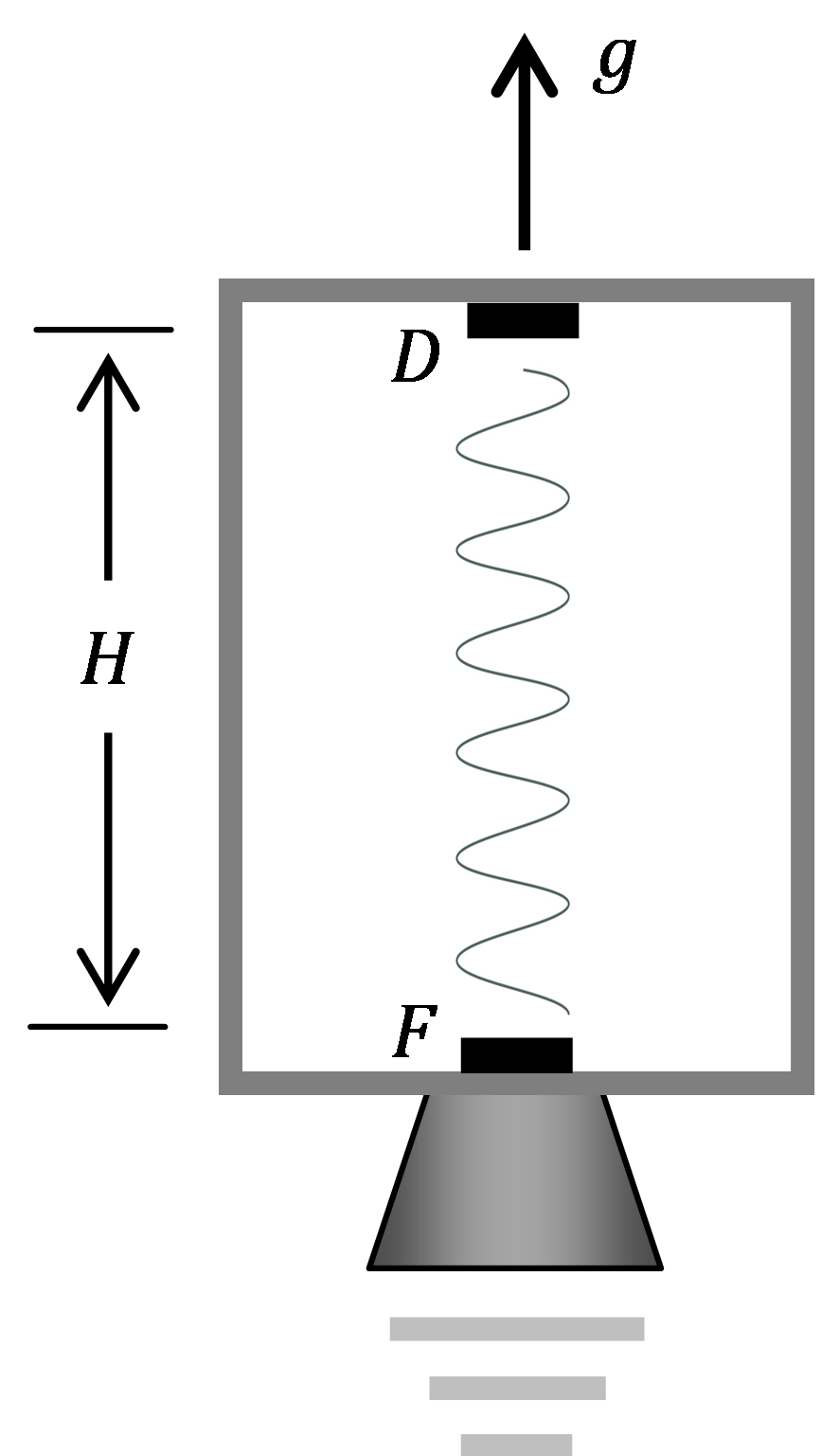}
  \caption{As the spacecraft moves with uniform acceleration $g$, the astronaut uses a detector $D$ to measure the wavelength of a photon emitted at $F$.}
\end{figure}

We can use the Einstein equivalence principle to predict the gravitational redshift of light and derive the equation that describes this effect. To do this, let us go back to the spacecraft described in Section 2 and suppose it is at rest in a region of space without gravity. On the floor of the spacecraft there is a light source $F$ that emits a photon of wavelength $\lambda_{0}$ vertically upwards. At the time of emission, the spacecraft begins to move with constant acceleration $g$. On the ceiling is a detector $D$ at a height $H$ (see Fig. 5).\\

As the spacecraft moves away from the emission point, a Doppler effect will occur, that is, $D$ will perceive that the photon wavelength has increased to a value $\lambda > \lambda_{0}$. As a first approximation, we can determine the relationship between $\lambda$ and $\lambda_{0}$ using the equation for the non-relativistic Doppler effect:

%eq 16
\begin{equation}
\frac{\lambda}{\lambda}_{0} = 1+ \frac{v}{c},
\end{equation}

where $v$ is the speed of the spacecraft when $D$ detects the photon. We can estimate the time it takes for the photon to reach $D$ as:

%eq 17
\begin{equation}
t = \frac{H}{c}.
\end{equation}

Since the spacecraft is initially at rest, its speed when $D$ detects the photon is $v = gt$, so that:

%eq 18
\begin{equation}
v = g \frac{H}{c}.
\end{equation}

By introducing this value of $v$ in Eq. (16), we obtain:

%eq 19
\begin{equation}
\frac{\lambda}{\lambda}_{0} = 1+ \frac{gH}{c^{2}}.
\end{equation}

We define redshift $z$ as the fractional change in wavelength:

%eq 20
\begin{equation}
z = \frac{\lambda}{\lambda_{0}}-1 = \frac{\lambda - \lambda_{0}}{\lambda_{0}} = \frac{\Delta \lambda}{\lambda_{0}} = \frac{gH}{c^{2}}.
\end{equation}

By virtue of the Einstein equivalence principle, we conclude that if the photon undergoes redshift within the spacecraft with acceleration $g$, it will also undergo offset if the spacecraft is at rest on Earth's surface (or on the surface of any celestial body), where the acceleration due to gravity is $–g$ (see Fig. 6). Then, Eqs. (19) and (20) must also be valid in a gravitational field.\\

\begin{figure}
  \centering
    \includegraphics[width=0.3\textwidth]{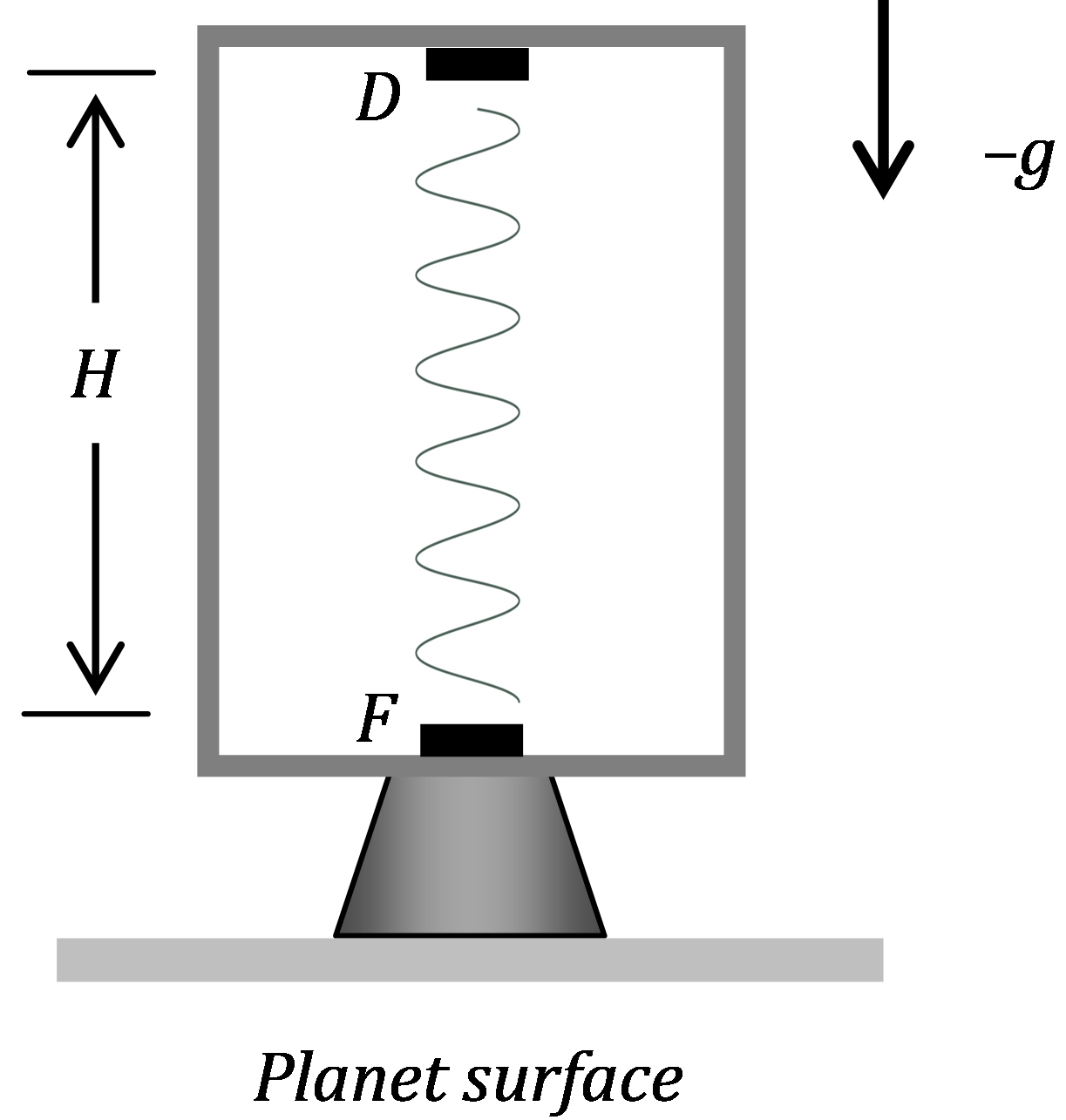}
  \caption{Spacecraft at rest on Earth's surface. A photon traveling towards $D$ undergoes a redshift analogous to that illustrated in Fig. 5.}
\end{figure}

Einstein's equation for gravitational redshift is equivalent to Eq. (19), which is a valid approximation when $H$ is very small compared to the Earth's radius.\\

The first experimental confirmation of the gravitational redshift of light was obtained 40 years after the astronomical expedition that verified the deflection of light by the Sun. In 1959, Robert Pound and Glen Rebka conducted an experiment at Harvard University, where they observed the redshift from photons that rose to a height of $22.6 m$. If in Eq. (20), we take $H = 22.6 m$ and $g = 9.81 m\cdot s^{-2}$, it is found that $z = 2.46 \times 10^{-15}$, which within the experimental uncertainties, agrees very well with the value found by Pound and Rebka [4].

\section{Final comments}
After the 1919 astronomical expedition, scientific interest in GR rapidly waned, and over the following decades, physicists turned their attention to other topics. This can be explained mainly by two factors: (1) GR is a mathematically very complex theory, and in Einstein's time only an exact solution of astronomical interest was known to the equations of GR (Schwarzschild solution), thus, there was little incentive to investigate the subject, since it seemed difficult to find new solutions of interest; (2) The empirical corroboration of GR was extremely difficult with the technological resources of that time.\\

Fortunately, in the 1960s the situation changed dramatically and GR experienced a renaissance. This is the time that Kip Thorne calls the golden age [5], during which new mathematical techniques were developed that made calculations easier. In addition, technology had advanced enough to allow accurate tests of GR to be carried out, such as the Glen and Rebka experiment, or other tests carried out shortly thereafter, such as the confirmation of the Shapiro time delay effect or the Hafele-Keating experiment. Furthermore, astronomical observations were beginning to reveal extreme phenomena that could only be adequately explained in the framework of GR (such as pulsars and quasars), so that physicists and astronomers began to pay attention to Einstein's theory. The reader who wants to delve into these topics can turn to the excellent Thorne’s book mentioned earlier. Another highly recommended popular science book is [3].\\

With ups and downs, since the golden age, interest in GR has not waned. In fact, over the past few years, interest has boomed, where new astronomical observations have made GR take on an importance it probably never had before, even in the golden age. Two milestones in this regard are the detection of gravitational waves in 2015 and the first picture of a black hole, obtained in 2019. In both cases, and in many others that we cannot analyse here, GR has been successfully confirmed. Everything seems to indicate that this is just the beginning of a period where GR will experience a second golden age. In this scenario, it is to be hoped that looking to the past will contribute to making physics teachers and students more prepared to understand and value current and future advances in GR.

\section*{Acknowledgments}
I would like to thank to Daniela Balieiro for their valuable comments in the writing of this paper. 

\section*{References}

[1] J. Pinochet, Einstein ring: Weighing a star with light, Phys. Educ., 53 055003 (2018).

\vspace{2mm}

[2] K. Krane, Modern Physics, 3 ed., John Wiley and Sons, Hoboken, 2012.

\vspace{2mm}

[3] C.M. Will, Was Einstein Right?, Basic Books, New York, 1986.

\vspace{2mm}

[4] R.V. Pound, G.A. Rebka, Gravitational red-shift in nuclear resonance, Phys. Rev. Lett., 3 (1959) 439-441.

\vspace{2mm}

[5] K.S. Thorne, Black Holes and Time Warps: Einstein's Outrageous Legacy, Norton, W. W. and Company, Inc., New York, 2014.

\end{document}